\documentclass[letterpaper, 10 pt, conference]{ieeeconf}
\IEEEoverridecommandlockouts                              
\usepackage[scaled=0.92]{helvet}
\usepackage{courier}
\usepackage{xspace}
\usepackage{color}
\usepackage{amsmath,amssymb,amsfonts}
\usepackage{mathrsfs}
\usepackage{graphicx}
\usepackage{subfigure}
\usepackage{array,makecell}

\usepackage{tikz}
\usepackage{xspace}

\DeclareRobustCommand{\legendline}[1]{\hspace{-3pt}
\tikz[#1,line width=0.4mm,baseline=-0.5ex]{\draw (0,0) -- (.35,0);}
\hspace{-3pt}}

\definecolor{mblue}{rgb}{0,0.4470,0.7410}
\definecolor{morange}{rgb}{0.8500,0.3250,0.0980}
\definecolor{myellow}{rgb}{0.9290,0.6940,0.1250}
\definecolor{mpurple}{rgb}{0.4940,0.1840,0.5560}
\definecolor{mgreen}{rgb}{0.4660,0.6740,0.1880}
\definecolor{mcyan}{rgb}{0.3010,0.7450,0.9330}
\definecolor{mred}{rgb}{0.6350,0.0780,0.1840}
\definecolor{mgreenblue}{rgb}{0.0,1.0,0.5}
\definecolor{parulablue}{rgb}{0.2431,0.1490,0.6588}
\definecolor{parulalblue}{RGB}{39,151,235}
\definecolor{parulagreen}{RGB}{129,204,89}
\definecolor{parulayellow}{RGB}{249,251,21}

\newcommand{\norm}[1]{\left\lVert#1\right\rVert}

\newcommand{\ltwo}{\ensuremath{\mathcal{L}_2}\xspace}

\newcommand{\m}[1]{\mathcal{#1}}
\newcommand{\mr}[1]{\mathrm{#1}}

\DeclareMathOperator{\col}{col}

\usepackage[noadjust]{cite}
\newcommand{\diag}[1]{\mathrm{diag}(#1)}
\newcommand{\vect}[1]{\mathrm{vec}\left(#1\right)}

\title{\LARGE \bf
Scheduling Dimension Reduction of LPV Models - A Deep Neural Network Approach 
}

\author{P.J.W. Koelewijn and R. T\'oth
\thanks{P.J.W. Koelewijn and R. T\'oth are with Control System Group, Faculty of  Electrical Engineering, Eindhoven University of Technology, 5600 MB Eindhoven, The Netherlands
        {\tt\small $\lbrace$p.j.w.koelewijn, r.toth$\rbrace$@tue.nl}}%
\thanks{This work has received funding from the European Research Council (ERC) under the European Union’s Horizon 2020 research and innovation programme (grant agreement nr. 714663).}%
}

\begin{document}

\maketitle
\thispagestyle{empty}
\pagestyle{empty}

\begin{abstract}
In this paper, the existing Scheduling Dimension Reduction (SDR) methods for Linear Parameter-Varying (LPV) models are reviewed and a Deep Neural Network (DNN) approach is developed that achieves higher model accuracy under scheduling dimension reduction. The proposed DNN method and existing SDR methods are compared on a two-link robotic manipulator, both in terms of model accuracy and performance of controllers synthesized with the reduced models. The methods compared include SDR for state-space models using Principal Component Analysis (PCA), Kernel PCA (KPCA) and Autoencoders (AE). On the robotic manipulator example, the DNN method achieves improved representation of the matrix variations of the original LPV model in terms of the Frobenius norm compared to the current methods. Moreover, when the resulting model is used to accommodate synthesis, improved closed-loop performance is obtained compared to the current methods.
\end{abstract}

\section{Introduction}

Over the last few decades, the Linear Parameter-Varying (LPV) framework has been developed into a powerful tool for both analysis and controller synthesis for nonlinear (NL) systems, see \cite{Hoffmann2015ASO} for an overview. Whereas most tools for analysis and synthesis have matured, the embedding of NL systems in LPV representations is still underdeveloped and depends on the expertise of the user. Some automated procedures exists, see e.g. \cite{Kwiatkowski2006AutomatedGA,Toth2010modeling,Casella2009automatic,Shamma1993gain}, but it is generally true that these procedures and heuristic methods can easily lead to an LPV model with a large amount of scheduling-variables. This is an issue when the LPV model is used for synthesis, due to the complexity of the synthesis procedure being proportional to the number of scheduling-variables. Limiting the number of scheduling-variables in the LPV embedding is therefore highly important for tractable LPV controller synthesis. 
Moreover, during the embedding procedure, control objectives are not taken into account, hence the LPV embedding can be conservative for the required control objectives. Hence, the objective of Scheduling Dimension Reduction (SDR) is to reduce this conservativeness, which can be achieved by taking into account a set of scheduling-variable trajectories associated with the expected closed-loop behavior of the system.  In literature, based on this concept, several data-based methods have been proposed to reduce the scheduling dimension of LPV models. 
These methods only focus on the reduction of the number of scheduling-variables and not on the reduction of the amount of states (referred to as model reduction), for which also several methods exist, see e.g. \cite{Toth2012OnTS}. Methods for reducing the amount of scheduling-variables based on data include methods based on Principal Component Analysis (PCA) \cite{Kwiatkowski2008PCABasedPS}, Kernel PCA (KPCA) \cite{Rizvi2016AKP} and Autoencoders (AE) \cite{Rizvi2018ModelRI}. 
While KPCA and the AE method use nonlinear mappings to construct the new scheduling-variable, showing improvement over the PCA method, they require an extra optimization step in order to synthesize an inverse transformation that enables the realization of the reduced LPV state-space model with affine dependency. Therefore, as a contribution of this paper, a novel method is developed, which like the AE method, proposed in \cite{Rizvi2018ModelRI}, uses a Neural Network (NN) in order to perform SDR, but from which the matrices of the reduced LPV state-space model can directly be extracted. This avoids the use of a second optimization step, hence, leads to more optimal results. Moreover, addition of hidden layers in the encoding layer is proposed in order to allow to handle more complex scheduling mappings. This paper compares the developed Deep Neural Network (DNN) approach to the existing methods when the SDR methods are applied to a two-link robotic manipulator. The latter also represents a contribution of the paper as the existing PCA, KPCA and AE methods have not been thoroughly compared in the literature.
Other methods for dimensionality reduction for LPV models, not relying on data, exist, where the LPV model is represented by a Linear Fractional Representation, see e.g. \cite{Varga1998automated,HECKER2005523,beck2006coprime}. These methods were not considered in this work as they aim only to the reduction of the extracted $\Delta$-block using a controllability/observability argument, not the reduction of the number of scheduling-variables.

The paper is structured as follows, in Section \ref{sec:problem}, a mathematical problem definition of the SDR problem will be given. Section \ref{sec:methods} gives an overview of the existing SDR methods. In Section \ref{sec:DNN}, the developed DNN approach is explained. A comparison of the DNN and existing methods is given in Section \ref{sec:results} on the example of a two-link planar robot manipulator both in terms of the modeling error and achieved performance of the controller that is synthesized based on the reduced model. Finally, in Section \ref{sec:conclusion}, conclusions on the given results are drawn.

\subsubsection*{Notation}
The Frobenius norm of a matrix $A \in \mathbb{R}^{n \times m}$ is denoted by $\lVert{A}\rVert_\mathrm{F}=\sqrt{\mathrm{trace}(AA^\top)}$.
The identity matrix of size $N$ is denoted by $I_N$. $A = \diag{A_1,\, \dots,\, A_n}$ denotes the block diagonal matrix $A$ with matrices $A_1$ to $A_n$ on the diagonal. For the matrix $M\in \mathbb{R}^{m\times n}$, $\vect{M}\in\mathbb{R}^{m\cdot n}$ denotes the vectorized form of the matrix $M$. The set $\left\{1,\, 2,\, \dots,\, N\right\}$ is denoted by $\mathbb{I}_1^N$. For a matrix $X$, $X_{i,j}$ denotes its element in the $i$'th row and $j$'th column, with $X_{i,*}$ and $X_{*,i}$ denoting the full $i$'th row and column, respectively. The notation $\col(x_1,\,\dots,\,x_n)$ denotes the column vector $\begin{bmatrix}
	x_1^\top & \cdots & x_n^\top
\end{bmatrix}^\top$.

\section{Problem Definition}\label{sec:problem}
\subsection{LPV Embedding}
Consider a NL dynamical system with a state-space representation given by 
\begin{equation}\label{eq:nonlinsys}
\begin{aligned}
\dot{x}(t) &= f\left(x(t),u(t)\right);\\
y(t) &= h\left(x(t),u(t)\right);
\end{aligned}
\end{equation} 
where $x(t) \in \mathbb{R}^{n_\mathrm{x}}$ is the state variable associated with the considered state-space representation of the system, $u(t) \in \mathbb{R}^{n_\mathrm{u}}$ is the input, and $y(t) \in \mathbb{R}^{n_\mathrm{y}}$ is the output of the system. $f$ and $h$ are defined as functions $\mathbb{R}^{n_\mathrm{x}} \times \mathbb{R}^{n_\mathrm{u}}\rightarrow\mathbb{R}^{n_\mathrm{x}}$ and $\mathbb{R}^{n_\mathrm{x}} \times \mathbb{R}^{n_\mathrm{u}}\rightarrow\mathbb{R}^{n_\mathrm{y}}$ respectively, and assumed to be Lipschitz continuous. As considered in the LPV literature, embedding of the NL system \eqref{eq:nonlinsys} in an LPV representation, corresponds to constructing
\begin{equation}\label{eq:lpvsys}
	\begin{aligned}
	\dot{x}(t) &= A(\rho(t)) x(t)+B(\rho(t))u(t);\\
	y(t) &= C(\rho(t)) x(t)+D(\rho(t))u(t);
 \end{aligned}
\end{equation}
where $\rho(t) \in \mathbb{R}^{n_{\rho}}$ is the scheduling-variable and there exists a function $\eta: \mathbb{R}^{n_\mathrm{x}} \times \mathbb{R}^{n_\mathrm{u}} \rightarrow \mathbb{R}^{n_\rho}$, called the scheduling map, such that $\eta(x(t),u(t)) = \rho(t)$. This proxy representation of \eqref{eq:nonlinsys} is used during control synthesis by confining $\rho(t)$ into a compact convex set $P \subset \mathbb{R}^{n_{\rho}}$ and synthesizing a controller that ensures stability and performance of \eqref{eq:lpvsys} under all possible variations of $\rho(t)\in P$. Therefore the embedding is constructed on the compact sets $X$, $U$, where $x\in X$ and $u\in U$, such that $P\supseteq\eta(X,U)$, often taken as the convex hull of $\eta(X,U)$. Moreover, it is assumed that the embedding  of \eqref{eq:nonlinsys} is performed such that the LPV system has an affine scheduling dependency, meaning that 
\begin{equation}\label{eq:affstruct}
	M(\rho) = \begin{bmatrix}
		A(\rho) & B(\rho)\\ C(\rho) & D(\rho)
	\end{bmatrix} = \underbrace{\begin{bmatrix}
		A_0 & B_0\\ C_0 & D_0
	\end{bmatrix}}_{M_0}+\sum_{i=1}^{n_\rho} \underbrace{\begin{bmatrix}
		A_i & B_i\\ C_i & D_i
	\end{bmatrix}}_{M_i} \rho_i,
\end{equation}
where $\rho = \col(\rho_1,\,\dots,\,\rho_{n_\rho})$ and $M_i \in \mathbb{R}^{m\times n},\, \forall \,i \in \mathbb{I}_1^{n_\rho}$ with $m=n_\mathrm{x}+n_\mathrm{y}$, $n = n_\mathrm{x}+n_\mathrm{u}$.

\subsection{Scheduling Dimension Reduction Problem}
By restricting the scheduling dependency to being affine, the scheduling map $\eta$ can contain many nonlinear functions that are dependent on the same elements of $x$ and $u$. Hence, the variation of scheduling-variables are not independent from each other, which contributes to the conservativeness of the LPV model \cite{Shamma1992gain}. This can have dramatic effects on the capability of the LPV synthesis to find a stabilizing controller for \eqref{eq:nonlinsys} via \eqref{eq:lpvsys} with acceptable performance. Therefore, reducing conservativeness through SDR can help with attaining performance requirements. Given an LPV embedding \eqref{eq:lpvsys} of \eqref{eq:nonlinsys} and a set of nominal scheduling-variable trajectories of \eqref{eq:lpvsys}, denoted by $\mathcal{D}$ that correspond to typical expected behavior of \eqref{eq:nonlinsys} through $\eta$, find an LPV embedding given by
\begin{equation}\label{eq:lpvsysred}
	\begin{aligned}
	\dot{x}(t) &= \hat{A}(\phi(t)) x(t)+\hat{B}(\phi(t))u(t);\\
	y(t) &= \hat{C}(\phi(t)) x(t)+\hat{D}(\phi(t))u(t);
 \end{aligned}
\end{equation}
which approximates \eqref{eq:lpvsys} under all trajectories of $\rho \in \m{D}$ and where $\phi(t) \in \Phi \subset \mathbb{R}^{n_{\phi}}$ is the (reduced) scheduling-variable with $n_\phi \leq n_\rho$ and \eqref{eq:lpvsysred} has an affine parameter dependency, i.e. a structure like \eqref{eq:affstruct}, with $\phi = \col(\phi_1,\, \cdots,\,\phi_{n_\phi})$, where $\Phi$ is compact convex set. Here, approximation is considered in the sense that $\phi = \mu \circ \eta $, with $\mu: \mathbb{R}^{n_\rho} \rightarrow \mathbb{R}^{n_\phi}$, where $\mu(\rho(t)) = \phi(t)$ is chosen such that \begin{equation}\label{eq:matdep}
	\hat{M}(\phi) = \begin{bmatrix}
		\hat{A}(\phi) & \hat{B}(\phi)\\ \hat{C}(\phi) & \hat{D}(\phi)
	\end{bmatrix} \approx M({\rho}).
\end{equation}
If possible, also find the inverse mapping $\mu^{-1}$, $\mu^{-1}(\phi(t)) = \hat{\rho}(t)$, where $\hat{\rho}(t)$ is an approximation of the original scheduling-variable $\rho(t)$, such that $\hat{M}(\phi) = M(\hat{\rho})$. The SDR problem is solved if \eqref{eq:lpvsysred} is a satisfactory approximation of \eqref{eq:lpvsys}. 
A satisfactory approximation is achieved, e.g., when, given a user defined $\varepsilon$, $\lVert M(\rho)-\hat{M}(\mu(\rho))\rVert_\mathrm{F}<\varepsilon$ for all $\rho \in \mathcal{D}$.
\section{Overview of SDR Techniques}\label{sec:methods}
In this section we briefly explain the procedures of the existing data based SDR methods in the literature.
\subsection{Preliminaries}
Assume an NL system given by \eqref{eq:nonlinsys}, embedded in an LPV representation \eqref{eq:lpvsys}. Furthermore, assume that we have generated a set of nominal trajectories of the scheduling-variable $\rho(t) = \eta(x(t),u(t))$, based on the trajectories of the NL system, sampled at time instances $t=k T_\mathrm{s}$, $k = 0,\,\dots,\,N-1$, given by \begin{equation}
	\Gamma = \begin{bmatrix}\label{eq:datamat}
		\rho\left(0\right) & \cdots & \rho\left((N-1)T_\mathrm{s}\right)
	\end{bmatrix},
\end{equation}
with $\Gamma \in \mathbb{R}^{n_\rho \times N}$ and $\mathcal{D} = \lbrace \rho(k T_\mathrm{s})\rbrace_{k=0}^{N-1}$. Moreover we introduce the shorthand notation $\rho_{(i)}:=\rho((i-1)T_\mr{s}) = \Gamma_{*,i}$.
The trajectories in $\Gamma$ are then normalized by an affine function $\mathcal{N}$,
 such that each row of the data matrix varies in $[-1,1]$, resulting in a normalized data matrix
	$\Gamma_\mathrm{n} = \mathcal{N}(\Gamma)$.
For all the discussed algorithms (also the DNN method in Section \ref{sec:DNN}) it is assumed\footnote{While not explicitly written down for the discussed methods, the normalization transformation does have to be taken into account for the mapping $\mu$ (and $\mu^{-1}$) and during construction of $\hat{M}(\phi)$.} that the data matrix, and therefore also $\rho_{(i)}$, is normalized, meaning that $\Gamma \equiv \Gamma_\mr{n}$. Next, the algorithms of the considered SDR methods will be discussed.

\subsection{Principal Component Analysis}\label{sec:PCA}
One of the earliest works to perform SDR based on trajectory data makes use of PCA \cite{Kwiatkowski2008PCABasedPS}.
The core idea of the PCA method is to extract the most significant directions, principal components, of the scheduling data $\Gamma$. For the PCA method, an SVD is done on $\Gamma$ in order to obtain the principal components, such that
\begin{equation}
	\Gamma = \begin{bmatrix}U_\mathrm{s} & U_\mathrm{r}\end{bmatrix}\begin{bmatrix}
		\Sigma_\mathrm{s} & 0 & 0\\0 & \Sigma_\mathrm{r} & 0
	\end{bmatrix}\begin{bmatrix}V_\mathrm{s}^\top \\ V_\mathrm{r}^\top\end{bmatrix},
\end{equation}
where $U_\mathrm{s} \in \mathbb{R}^{n_\rho \times n_\phi}$, $V_\mathrm{s} \in \mathbb{R}^{N\times n_\phi}$ are unitary and $\Sigma_\mathrm{s} \in \mathbb{R}^{n_\phi\times n_\phi}$ is positive diagonal, containing the highest $n_\phi$ singular values of $\Gamma$. $n_\phi$ can be chosen by the user to govern the complexity of the reduced model. Then, an approximation of the data matrix, denoted by $\hat{\Gamma}$, is given by $\hat{\Gamma} = U_\mathrm{s} \Sigma_\mathrm{s} V_\mathrm{s}^\top \approx \Gamma$.
The new reduced scheduling variable $\phi$ is then given by\begin{equation}
	\phi(t) = \mu(\rho(t)) = U_\mathrm{s}^\top \rho(t),
\end{equation}
and an approximation of the old scheduling variable $\rho$ is obtained via\vspace{-.5em}
\begin{equation}\label{eq:pcainvmap}
	\hat{\rho}(t) = \mu^{-1}(\phi(t)) = U_\mathrm{s}\phi(t).
\end{equation}
The matrices of the (scheduling dimension) reduced LPV system can then be constructed by using the relation $\hat{M}(\phi) = M(\hat{\rho})$ and \eqref{eq:pcainvmap}. 
For a more in-depth overview of the PCA method for SDR see \cite{Kwiatkowski2008PCABasedPS}.


\subsection{Kernel PCA}\label{sec:KPCA}
The next method discussed is the kernel PCA method \cite{Rizvi2016AKP}. A disadvantage of the PCA 
method is that it only seeks a linear mapping for $\mu$. The KPCA method extends the PCA method by allowing a nonlinear mapping. In the KPCA method, the data is nonlinearly mapped to a higher dimensional, so called, feature space on which normal PCA is applied. This mapping into the feature space is denoted by $\Theta$. The covariance matrix is then given by
\begin{equation}
	\bar{C}:=\frac{1}{N} \sum_{j=1}^{N} \Theta\left(\rho_{(j)}\right) \Theta^{\top}\hspace{-3pt}\left(\rho_{(j)}\right).
\end{equation}
The principal components are then computed, such that $\lambda_l v_l = \bar{C}v_l$ holds, where $\lambda_l$ is the $l$'th eigenvalue and $v_l$ is the $l$'th eigenvector.
By this method, we seek for a mapping to an appropriate feature space where PCA will result in the smallest number of components. The mapping $\Theta$ is a priori not known. In KPCA, the idea is to characterize the inner product of $\Theta$ with an a priori chosen kernel function, resulting in the kernel matrix
\begin{equation}\label{eq:kernelmat}
	K_{ij} = (\Theta(\rho_{(i)})^\top \Theta(\rho_{(j)})) = k(\rho_{(i)},\rho_{(j)}),
\end{equation}
where $K \in \mathbb{R}^{N\times N}$ and $k(\cdot , \cdot)$ is a nonlinear kernel function. Examples of the kernel function include, the sigmoid kernel $k(\rho_{(i)},\rho_{(j)}) = \tanh(\kappa(\rho_{(i)}^\top \rho_{(j)})+\iota)$, the radial basis function $k(\rho_{(i)},\rho_{(j)}) = \exp(-\norm{\rho_{(i)}-\rho_{(j)})}^2/\kappa^2)$ and polynomial kernel $k(\rho_{(i)},\rho_{(j)}) = (\rho_{(i)}^\top \rho_{(j)}+\iota)^\kappa$, where $\kappa$ and $\iota$ are hyperparameters chosen such that PCA can be accomplished with the lowest number of components given the structure of $k$. The data in the feature space is assumed to be centered, which is not always the case, therefore the centered kernel matrix is constructed as follows
\begin{equation}
	K_\mathrm{c} = K-1_\mathrm{N} K - K 1_\mathrm{N} + 1_\mathrm{N} K 1_\mathrm{N},
\end{equation}
where $1_\mathrm{N} \in \mathbb{R}^{N\times N}$ denotes the matrix with each element being $\frac{1}{N}$. The principal components of $K_\mathrm{c}$ are then computed instead of $\bar{C}$. Resulting, for non-zero eigenvalues, in
\begin{equation}
	\lambda_l \alpha_l = K_\mathrm{c} \alpha_l,
\end{equation}
where $\alpha_l = \col(
	\alpha_{l,1}\, \dots \,\alpha_{l,N})$, with $v_l = \sum_{i=1}^N \alpha_{l,i} \Theta(\rho_{(i)})$, where each $\alpha_l$ (for $l \in \mathbb{I}_1^N$) is normalized by $1/\sqrt{\lambda_l}$. The new scheduling variable $\phi$ is then
\begin{equation}
	\phi_l(t) = \sum_{i=1}^N \alpha_{l,i} k(\rho_{(i)},\rho(t)) \quad \text{for}\; l \in \mathbb{I}_1^{n_\phi}.
\end{equation}
The inverse mapping $\mu^{-1}$ for KPCA cannot be analytically constructed in general and hence a further optimization step is required. We refer the reader to \cite{Rizvi2014ParameterSU} for the details. Due to the difficulty constructing the inverse mapping and due to this mapping being nonlinear, the affine scheduling-variable dependent matrices $\hat{A},\, \dots,\,\hat{D}$ are constructed by solving the following optimization problem
\begin{equation}\label{eq:matrixopt}
	\min_{\left\lbrace \hat{M}_i\right\rbrace_{i=1}^{n_\phi}} \quad \mathcal{I} = \frac{1}{N}\sum_{j=1}^N \norm{M\left(\rho_{(j)}\right)-\hat{M}\left(\phi_{(j)}\right)}_\mathrm{F}^2,
\end{equation}
where $\phi_{(j)} := \phi((j-1)T_\mathrm{s})$.
See \cite{Rizvi2016AKP} for a more in-depth overview of the KPCA method for SDR.


\subsection{Autoencoder}\label{sec:AE}
The final method that will be discussed is the AE method \cite{Rizvi2018ModelRI}. The AE  method like KPCA uses nonlinear functions for the mapping $\mu$. This method makes use of an AE NN, which is trained on the data set to construct the mappings $\mu$ and $\mu^{-1}$. Therefore it has as an advantage over KPCA that $\mu^{-1}$ is co-synthesized with $\mu$. Again, assume we have an LPV system given by \eqref{eq:lpvsys}, and a data matrix given by \eqref{eq:datamat}. The data matrix is fed into a two-layer\footnote{In general, the AE can have more than two layers, but as in \cite{Rizvi2018ModelRI} this is not considered for this method in this work.}
 AE NN,
 which consist of a single encoding layer (from $\rho$ to $\phi$) and a decoding layer (from $\phi$ to $\hat{\rho}$). On the first layer (the encoding layer), we have the following relation
\begin{equation}
\phi_{(i)}=h^{[1]}\left(W^{[1]} \rho_{(i)}+b^{[1]}\right) \quad \text{for}\; i \in \mathbb{I}_{1}^N,
\end{equation}
where $h$ is called the activation function, with the weight matrix $W^{[1]} \in \mathbb{R}^{n_\phi \times n_\rho}$, and bias vector $b^{[1]} \in \mathbb{R}^{n_\phi}$. Examples of activation functions include the logistic sigmoid (logsig) function and hyperbolic function (tanh). 

On the second layer, on the decoding side, we have the following relation
\begin{equation}\label{eq:aedecoding}
\hat{\rho}_{(i)}=h^{[2]}\left(W^{[2]} \phi_{(i)}+b^{[2]}\right)\quad \text{for}\; i \in \mathbb{I}_{1}^N,
\end{equation}
where $W^{[2]} \in \mathbb{R}^{n_\rho \times n_\phi}$ and $b^{[2]} \in \mathbb{R}^{n_\rho}$. To train the AE, an optimization procedure is used to minimize the error between $\rho_{(i)}$ and $\hat{\rho}_{(i)}$ for $i \in \mathbb{I}_1^N$. This is done by solving the following optimization problem
\begin{equation}\label{eq:nnopt}
\min_{W^{[k]}, b^{[k]},\,k=1,2} \quad \frac{1}{n} \sum_{i=1}^{n_\rho} \sum_{j=1}^{N}\left(\Gamma_{i,j}-\hat{\Gamma}_{i,j}\right)^{2}.
\end{equation}
Additional terms can be incorporated into the cost function in order to reduce overfitting such as weight and/or sparsity regularization, see \cite{Rizvi2018ModelRI} for more details.
After optimization of \eqref{eq:nnopt} the new scheduling variable can be expressed as
\begin{equation}
	\phi(t) = \mu(\rho(t))= h^{[1]}\left(W^{[1]} \rho(t)+b^{[1]}\right),
\end{equation}
with the inverse mapping
\begin{equation}\label{eq:aeinvmap}
	\hat{\rho}(t) = \mu^{-1}(\phi(t))= h^{[2]}\left(W^{[2]} \phi(t)+b^{[2]}\right).
\end{equation}
However, due to the nonlinear inverse mapping \eqref{eq:aeinvmap}, again a separate optimization procedure is required as for the KPCA method, see Section \ref{sec:KPCA}, specifically \eqref{eq:matrixopt}, to obtain the affine scheduling-variable dependent matrices $\hat{A},\, \dots,\,\hat{D}$ of the reduced model.
See \cite{Rizvi2018ModelRI} for a detailed explanation of the AE method for SDR.

\section{Deep NN Approach}\label{sec:DNN}
While the AE and KPCA approaches have the benefit of using nonlinear mappings for the reduction, they still require an extra optimization step in order to obtain the matrices of the reduced LPV state-space model. However, the impact of this second optimization step, characterizing the model approximation error, is not taken into account in the synthesis of $\phi(t)$. Hence, as the contribution of this paper, a new method is developed which, like the AE method, uses an NN in order to construct the mapping $\phi = \mu(\rho)$. However, unlike the AE method, a vectorized form of $\hat{M}(\phi)$ is used as output of the network in order to directly construct the matrices of the reduced LPV model. In this way, the synthesis of $\phi(t)$ and the state-space matrices associated with it are co-determined under optimal approximation of $M(\rho)$ on $\m{D}$. Furthermore, (multiple) hidden layers are used in the encoding layer of the network, which allows to capture more complex mappings. Assume we have an LPV system given by \eqref{eq:lpvsys} and a data matrix given by \eqref{eq:datamat}, for this method called the input data matrix. Moreover assume we have another data matrix, the output data matrix, given by
\begin{equation}
	\Lambda = \begin{bmatrix}
		M_\mr{v}(\rho_{(1)}) & \cdots & M_\mr{v}(\rho_{(N)})
	\end{bmatrix},
\end{equation}
where $\Lambda\in \mathbb{R}^{\nu\times N}$, with $\nu = m\cdot n$, and where
\begin{equation}
	M_\mathrm{v}(\rho) = \vect{\sum_{i=1}^{n_\rho}M_i(\rho)} = \sum_{i=1}^{n_\rho} \vect{M_i} \rho_i.
\end{equation}

\begin{figure}
	\centering
	\vspace{-.5em}
	\includegraphics[scale=.85]{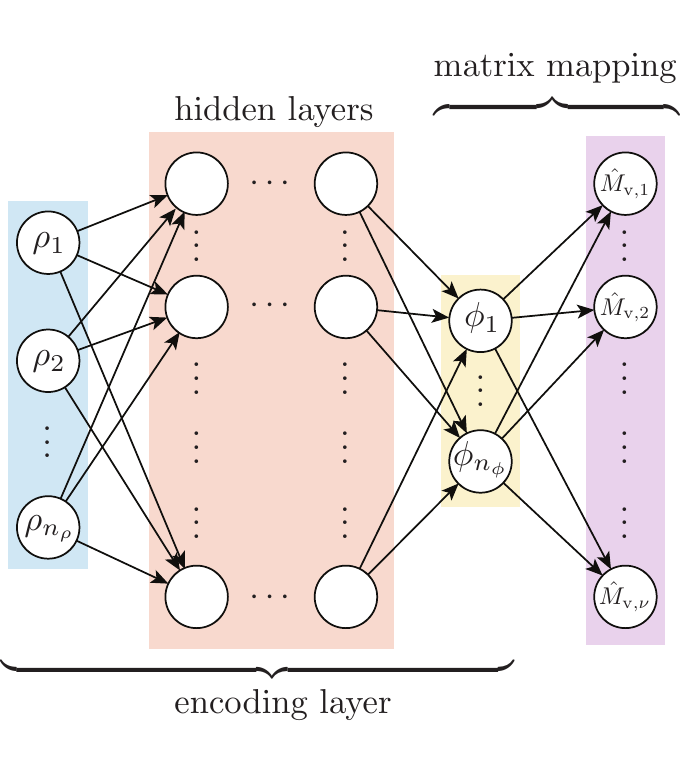}
	\vspace{-1.5em}
	\caption{DNN architecture.}
	\vspace{-1em}
	\label{fig:nn}
\end{figure}

The input data matrix is fed into the input layer of the NN, for which holds that
\begin{equation}\label{eq:nnhidlayer1}
	l^{[1]}_{(i)} = h^{[1]}\left(W^{[1]}\rho_{(i)}+b^{[1]}\right)\quad \text{for}\; i \in \mathbb{I}_{1}^N,
\end{equation}
where $l^{[1]} \in \mathbb{R}^{n_{l,1}}$ is the `output' of the first hidden layer, with the weight matrix $W^{[1]} \in \mathbb{R}^{n_{l,1} \times n_\rho}$, and bias vector $b^{[1]} \in \mathbb{R}^{n_{l,1}}$. For the second hidden layer till the $(g-1)$'th hidden layer (where $g$ denotes the number of hidden layers), we have the relation
\begin{equation}\label{eq:nnhidlayerk}
	l^{[k]}_{(i)} = h^{[k]}\left(W^{[k]}l^{[k-1]}_{(i)}+b^{[k]}\right)\; \text{for}\; k \in \mathbb{I}_{1}^{g-1}, \;\text{and}\; i \in \mathbb{I}_{1}^N,
\end{equation}
where $l^{[k]} \in \mathbb{R}^{n_{l,k}}$ is the `output' of the $k$'th hidden layer, with the weight matrix $W^{[k]} \in \mathbb{R}^{n_{l,k} \times n_{l,k-1}}$, and bias vector $b^{[k]} \in \mathbb{R}^{n_{l,k}}$. For the $g$'th hidden layer, i.e. the final hidden layer, we have
\begin{equation}\label{eq:nnhidlayerg}
	\phi_{(i)} = h^{[g]}\left(W^{[g]}l^{[g-1]}_{(i)}+b^{[g]}\right)\quad \text{for}\; i \in \mathbb{I}_{1}^N,
\end{equation}
with weight matrix $W^{[g]} \in \mathbb{R}^{n_\phi \times n_{\mr{n},g-1}}$, and bias vector $b^{[g]} \in \mathbb{R}^{n_\phi}$. Together, equations \eqref{eq:nnhidlayer1}-\eqref{eq:nnhidlayerg} make up the encoding layer of the NN and hence make up the mapping $\mu: \mathbb{R}^{n_{\rho}}\rightarrow \mathbb{R}^{n_{\phi}}$. While many different activation functions (as in the AE case) can be used in the encoding layer, we propose to use Rectified Linear Unit (ReLU) activation functions, given by $h(x)=\max(0,x)$. Deep neural networks using ReLU functions have been proven to be universal function approximators \cite{Hanin2017} with favorable benefits during training (stable non-vanishing gradient propagation). Moreover, ReLU functions have computational benefits, as their expression can be evaluated quickly and because they introduce sparsity in the network (as some neurons in a particular layer can be zero). Finally, we have the matrix mapping layer, which is the final layer in the network. Before giving the relation we first define the vector
\begin{equation}\label{eq:vecmhat}
	\hat{M}_\mathrm{v}(\phi) = \vect{\sum_{i=1}^{n_\phi}\hat{M}_i(\phi)} = \sum_{i=1}^{n_\phi} \vect{\hat{M}_i} \phi_i,
\end{equation}
where $\hat{M}_\mathrm{v}(\phi) = \col \left(
	\hat{M}_{\mr{v},1}(\phi),\, \hat{M}_{\mr{v},2}(\phi) ,\, \dots ,\, \hat{M}_{\mr{v},\nu}(\phi)
\right)\in \mathbb{R}^{\nu}$.
The final layer is then given by
\begin{equation}\label{eq:nnfinallayer}
	\hat{M}_\mr{v}(\phi_{(i)}) = W^{[g+1]}\phi_{(i)}+b^{[g+1]}\quad \text{for}\; i \in \mathbb{I}_{1}^N,
\end{equation}
where $W^{[g+1]}\in \mathbb{R}^{\nu \times n_\phi}$ with bias vector $b^{[g+1]} \in \mathbb{R}^{\nu}$. Note that the matrix mapping layer does \emph{not} use any activation functions, this is done such that the required affine relation between the reduced scheduling-variables and the (vectorized) reduced LPV system matrices is obtained. The encoding layer, \eqref{eq:nnhidlayer1}-\eqref{eq:nnhidlayerg}, together with the matrix mapping layer, \eqref{eq:nnfinallayer}, make up the full NN. The full NN structure is also depicted in Fig. \ref{fig:nn}.
To reconstruct the state-space matrices of the reduced LPV model, we can use \eqref{eq:vecmhat} and \eqref{eq:nnfinallayer} to obtain the relations
\begin{equation}
\begin{gathered}
	\vect{\hat{M}_0} = \vect{M_0}+ b^{[g+1]};\\
	\begin{bmatrix}\vect{\hat{M}_1} &\cdots & \vect{\hat{M}_{n_\phi}}\end{bmatrix} = W^{[g+1]}.
\end{gathered}
\end{equation}

The weights and biases of the layers of the DNN are trained by solving the following optimization problem
\begin{equation}\label{eq:nnopt}
\min_{W^{[k]}, b^{[k]},\,k=1,\dots,g+1} \quad \frac{1}{N}\sum_{j=1}^N \norm{M\left(\rho_{(j)}\right)-\hat{M}\left(\phi_{(j)}\right)}_\mathrm{F}^2,
\end{equation}
This optimization problem can then be solved by means of a backpropagation algorithm \cite{Goodfellow2016} in combination with a gradient decent algorithm, such as Stochastic Gradient Decent (SGD), Adam \cite{Kingma2014}, or AdaBound \cite{Luo2019}. Like in the AE method, in order to reduce overfitting, regularization techniques can be used such as regularization on the weights, sparsity regularization. Moreover, if multiple hidden layers are used in the encoding layer, regularization techniques such as dropout \cite{Srivastava2014} can be used to also reduce overfitting. Based on the above given considerations, implementation of \eqref{eq:nnopt} was developed in Python using Keras \cite{chollet2015keras}.  

\section{Results}\label{sec:results}
\begin{table}
\vspace{1em}
\caption{Physical parameters of robot manipulator.}
\label{tbl:param}
\centering
\begin{tabular}{l||c|c|c|c|c|c|c}
\textbf{Parameter:} & $a$      & $b$     & $c$      & $d$   & $e$   & $f$ & $n$ \\ \hline
\textbf{Value:}     & 5.6794 & 1.473 & 1.7985 & 0.4 & 0.4 & 2 & 1
\end{tabular}\vspace{-1.2em}
\end{table}
In this section the methods discussed from Section \ref{sec:methods} and Section \ref{sec:DNN} are compared on the LPV modeling and control design problem of a two-link planar robot manipulator \cite{Kwiatkowski2005LPVCO}.
\subsection{Robot Manipulator}
The robot manipulator can be described using the following equation of motion
\begin{equation}\label{eq:eom}
M(q(t)) \ddot{q}(t)+C(q(t), \dot{q}(t))+g(q(t))=n\tau(t),
\end{equation}
where $q(t) = \col(
	q_1(t),\, q_2(t) 
)$ are the angles, $\tau(t) = \col (\tau_1(t),\,\tau_2(t))$ the motor torques and
\begin{equation*} 
\begin{gathered}
M=\begin{bmatrix} {a} & {b \cos _{\Delta}(t)} \\ {b \cos _{\Delta}(t)} & {c}\end{bmatrix},
\quad g=\begin{bmatrix}{-d \sin \left(q_{1}(t)\right)} \\ {-e \sin \left(q_{2}(t)\right)}\end{bmatrix}, \\
C=\begin{bmatrix}{b \sin _{\Delta}(t) \dot{q}_{2}(t)^{2}+f \dot{q}_{1}(t)} \\ {-b \sin _{\Delta}(t) \dot{q}_{1}(t)^{2}+f\left(\dot{q}_{2}(t)-\dot{q}_{1}(t)\right)}\end{bmatrix},
\end{gathered}
\end{equation*}
where $\cos_\Delta(t) = \cos(q_1(t)-q_2(t))$ and $\sin_\Delta(t) = \sin(q_1(t)-q_2(t))$.
The values of the (physical) parameters of the robot manipulator are given in Table \ref{tbl:param}. The equations of motion \eqref{eq:eom} can be rewritten in a nonlinear state-space form \eqref{eq:nonlinsys}, resulting in
\begin{equation}\label{eq:robotss}
\begin{aligned}
	\dot{x}(t) &= \begin{bmatrix}
		x_2(t)\\M(x_1(t))^{-1}\left(n\tau(t)-C(x_1(t),x_2(t))+g(x_1(t))\right)
	\end{bmatrix};\\
	y(t) &= \begin{bmatrix} I_2 & 0 \end{bmatrix} x(t);
	\end{aligned}
\end{equation}
where $x(t) = \col(x_1(t), \, x_2(t))= \col( q(t),\, \dot{q}(t))$ and $u(t) = \col(\tau_1(t),\,\tau_2(t))$.

\subsection{LPV Model}
The LPV model of the robot manipulator, based on the LPV model in \cite{Kwiatkowski2005LPVCO}, is given by
\begin{equation}\label{eq:l2lvp}
\begin{aligned}
	\dot{x}(t) &= A(\rho(t))x(t)+B(\rho(t))u(t);\\
	y(t) &= Cx(t)+Du(t);
\end{aligned}
\end{equation}
where
\begin{equation}
\begin{aligned}
	A(\rho) &= \begin{bmatrix}
		0 & 0 & 1 & 0\\
		0 & 0 & 0 & 1\\
		c d \rho_3 & -b e \rho_4 & \rho_5 & b \rho_6\\
		- b d \rho_7 & a e \rho_8 & \rho_9 & \rho_{10}
	\end{bmatrix}, & C &= \begin{bmatrix} I_2 & 0\end{bmatrix},\\
	B(\rho) &= \begin{bmatrix}
		0&0\\0&0\\
		c n \rho_1 & -b n \rho_2\\-b n \rho_2 & a n \rho_1
	\end{bmatrix}, & D &= 0.
	\end{aligned}
\end{equation}
The scheduling map (only depending on the states) is $\eta: \mathbb{R}^4 \rightarrow \mathbb{R}^{10}$, which is given by
\begin{equation}\label{eq:l2map}
\eta(x) = 
	\left[\begin{array}{l}1/h\\
\cos_\Delta(t)/h\\
\mathrm{sinc}(x_1)/h\\
\cos_\Delta \mathrm{sinc}(x_2)/h\\
(-b^2 \sin_\Delta \cos_\Delta x_3-(c+b \cos_\Delta) f)/h\\
(-c \sin_\Delta x_4+\cos_\Delta f)/h\\
\cos_\Delta \mathrm{sinc}(x_1)/h\\
\mathrm{sinc}(x_2)/h\\
(a b \sin_\Delta x_3+f (a+b\cos_\Delta))/h\\
(b^2 \sin_\Delta \cos_\Delta x_4-af)/h
\end{array}\right],
\end{equation}
where $h = a c-b^2 (\cos_\Delta)^2$.
\subsection{Scheduling Dimension Reduction}
In order to perform SDR on \eqref{eq:l2lvp}, a data set is required containing a set of typical trajectories of the scheduling-variables. As the reduced LPV models will be used for controller synthesis and as the objective of the controller will be reference tracking, the to be followed reference trajectory is used as data set, see `Reference 1' in Fig. \ref{fig:track_nr1_l2}. To obtain the trajectories of the scheduling-variables, the mapping in \eqref{eq:l2map} is used. Note, that while not displayed in Fig. \ref{fig:track_nr1_l2}, the angular velocities corresponding to the reference trajectory are also required and used to obtain the data set. For the KPCA method, a sigmoid kernel is chosen with hyperparameters $\kappa=0.1$ and $\iota = 0.1$. For the AE method, logsig functions are used for both the decoding and encoding layer activation functions. Furthermore weight regularization is added with a coefficient of $1e{-5}$ and sparsity regularization is added with a coefficient of $0.5$. Moreover, for implementation and training of the AE based method the \texttt{Autoencoder} class from the MATLAB Deep Learning Toolbox was used. For the DNN approach, the AdaBound \cite{Luo2019} gradient descent method is used for training. For the encoding layer of the DNN approach, one hidden layer is used with 5 neurons and, as mentioned, ReLU functions are used as activation functions in the encoding layer. Moreover, for the DNN method, weight regularization was added with a coefficient of $1e{-6}$. The hyperparameters for each method were chosen such that their respective cost function were minimal.

\subsection{LPV Controller Design}
\begin{figure}
	\centering
	\includegraphics[scale=1]{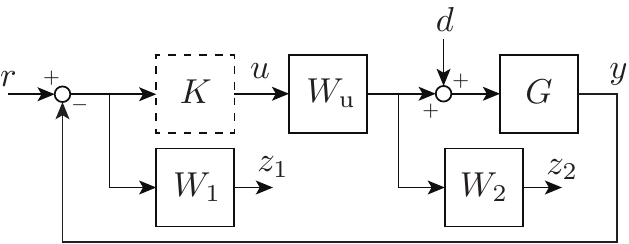}
	\caption{Generalized plant for controller synthesis.}
	\label{fig:genplant}
\end{figure}

For the controller design, a generalized plant is constructed, equivalent with the one used in \cite{Rizvi2016AKP}, in order to achieve reference tracking of $x_1$ and $x_2$ (i.e. the two angles, $q_1$ and $q_2$). The generalized plant is shown in Fig. \ref{fig:genplant}, where  $r$ is the reference signal and $d$ the input disturbance and $z_1$, $z_2$ are performance channels. An affine (\ltwo) LPV controller is synthesized, on the basis of the method in \cite{Apkarian1995}, which minimizes the $\ltwo$-gain from disturbance to performance channel. This synthesis procedure is then performed with the reduced plants ($G$ in Fig. \ref{fig:genplant}) resulting from the various SDR techniques. The resulting \ltwo-gains can be found in Table \ref{tbl:l2_gamma}. The weighting filter $W_1$ is chosen to include low-pass characteristics on both channels in order to ensure good tracking performance at low frequencies, while $W_2$ is chosen as a constant gain for both channels in order to limit the motor torques. The weighting filter $W_\mr{u}$ is chosen as a true low-pass filter on both channels\footnote{$W_\mr{u}$ is included because the synthesis procedure requires the relation from the control input to the state of the generalized plant to be independent of the scheduling-variable.}. The exact transfer functions applied in Fig. \ref{fig:genplant} are given in Table \ref{tbl:wfilt}\footnote{All the weighting filters are block diagonal with on the diagonal the same transfer function for both channels.}. 

\begin{table}
\centering
\caption{Weighting filters of the generalized plant where $s$ is the complex frequency.}
\label{tbl:wfilt}
\begin{tabular}{l||c|c|c}
\textbf{Weighting Filter:}   & $W_1(s)$                     & $W_2(s)$  & $W_\mathrm{u}(s)$      \\ \hline
\textbf{Transfer Function:} & $\frac{(5\cdot 10^{-1})s+5}{s+(5\cdot 10^{-5})}$ & $3\cdot 10^{-3}$ & $\frac{1\cdot 10^3}{s+(1\cdot 10^3)}$ 
\end{tabular}\vspace{-1em}
\end{table}

\subsection{Results}\label{sec:l2results}
\subsubsection{Modeling error}
\begin{figure}
	\centering
	\includegraphics[scale=1]{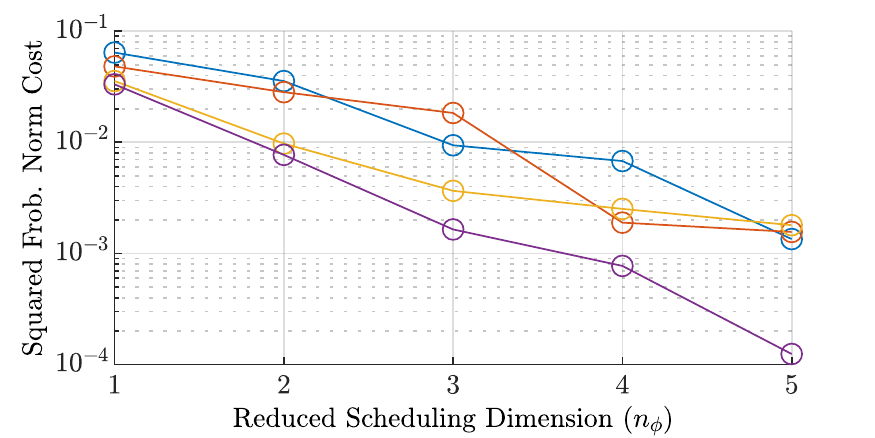}
	\caption{Average squared Frobenius Norm cost for different scheduling sizes ($n_\phi$) using PCA (\legendline{mblue}), KPCA (\legendline{morange}), AE (\legendline{myellow}), DNN (\legendline{mpurple}).}
	\label{fig:frobcost}
\end{figure}

The approximation error, in terms of average squared Frobenius norm (see \eqref{eq:matrixopt}), of the reduced LPV models is compared for various reductions to a scheduling size $n_\phi$. These results are given in Fig. \ref{fig:frobcost}.
From the results it can be concluded that the new developed DNN method results overall in the best performance for all the considered scheduling sizes. Due to directly optimizing the average squared Frobenius cost using the developed DNN method, instead of needing an extra optimization step as in the KPCA or AE method, a lower cost can be achieved. Moreover, the scaling to larger scheduling dimensions is also better with the developed DNN method compared to the other methods.

\subsubsection{LPV controller performance}
\begin{figure}
	\centering
	\vspace{1pt}
	\includegraphics[scale=1]{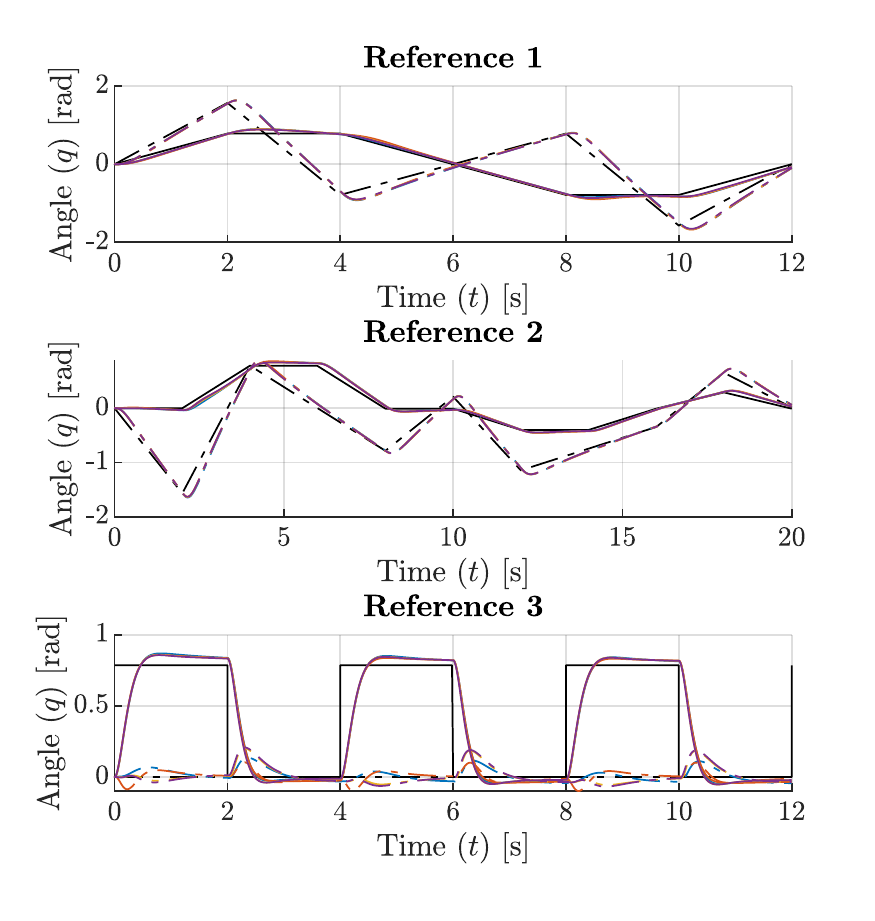}\vspace{-2em}
	\caption{Tracking performance of LPV controllers w.r.t. reference \protect\linebreak $q_1$ (\legendline{black}) and $q_2$ (\legendline{black,dashed}) using the SDR techniques with $n_\phi = 1$, using PCA (\legendline{mblue}), KPCA (\legendline{morange}), AE (\legendline{myellow}), DNN (\legendline{mpurple}).}
	\label{fig:track_nr1_l2}
\end{figure}

The controllers resulting from the synthesis on the reduced LPV models (with $n_\phi=1$) are interconnected with the nonlinear model of the robot arm and the tracking performance is evaluated.
The tracking performance of the various resulting controllers is evaluated for three reference trajectories. The first reference trajectory (reference 1) is the reference trajectory also used to construct the data set. The second reference trajectory (reference 2) is constructed to have similar `behavior' as reference 1, while still being different from the first reference.  Reference 2 is used to analyze if the methods overfit. Finally, the third reference trajectory (reference 3) consists of a square wave trajectory for $q_1$ and $q_2$ kept at zero. Reference trajectory 3 has different `behavior' compared to reference 1 and 2 (hence not present in the data used for SDR) and is used to analyze the robustness of the methods. 
\begin{table}
\vspace{-.5em}
\centering
\caption{\ltwo-gains obtained for the compared SDR methods for ${n_\phi = 1}$.}
\vspace{-1em}
\label{tbl:l2_gamma}
\begin{tabular}{l||c|c|c|c}
\textbf{SDR method:}       & PCA & KPCA & AE   & DNN \\ \hline\hline
\textbf{$\mathbf{\ltwo}$-gain:} & 1.25 & 1.26 & 1.25 & 1.24 \\ 
\end{tabular}
\vspace{-1.5em}
\end{table}
Shown in Fig. \ref{fig:track_nr1_l2} are the simulation results for the considered reference trajectories using the resulting controllers (i.e. using various SDR techniques). It is apparent from these figures that for this example the methods all result in trajectories that are close together. 
Therefore, the \ltwo-gains for the different controllers is also computed/approximated based on the simulation data, i.e.
\begin{equation}\label{eq:perfcomp}
\gamma = \frac{\sqrt{\int_0^\infty \vert z(t)\vert^2\, dt} }{\sqrt{\int_0^\infty \vert w(t)\vert^2\, dt}}\approx \frac{\sqrt{\sum_{k=0}^T\vert z(k T_\mathrm{s})\vert^2}}{\sqrt{\sum_{k=0}^T \vert w(k T_\mathrm{s})\vert^2}},
\end{equation}
where $z(t) = \col (z_1(t),\, z_2(t))$, $w(t) = \col(r(t),\,d(t))$ (see Fig. \ref{fig:genplant}) and $T$ is the end-time of the simulation. The \ltwo-gains based on the simulation data, computed using \eqref{eq:perfcomp}, for each of the different reference trajectories are given in Table \ref{tbl:l2_sim_perf}. From the computed \ltwo-gains it is clear that the DNN method provides overall the best controller performance. An additional benefit of the proposed DNN method is that due to the use of ReLU activation functions, the mapping $\mu$ from $\rho(t)$ to $\phi(t)$ can be computed faster compared to the AE method proposed in \cite{Rizvi2018ModelRI} and the KPCA methods, which use more complex nonlinear functions. This makes the DNN method also better suited for realtime implementation.


\begin{table}
\centering
\caption{\ltwo-gain performance of LPV controllers in simulation using the different SDR techniques with $n_\phi=1$.}
\label{tbl:l2_sim_perf}
\begin{tabular}{c||c|c|c|c}
       & PCA & KPCA & AE   & DNN \\ \hline\hline
Ref. 1 & 0.9466  &  1.008  &  0.9532   & \textbf{0.9062} \\
Ref. 2 & 0.9564  &  0.9109 &  0.9088   & \textbf{0.8834} \\
Ref. 3 & 2.044   &  \textbf{1.333} &   1.511  &  1.500
\end{tabular}
\vspace{-10pt}
\end{table}


\section{Conclusion}\label{sec:conclusion}
In this paper a novel SDR method, that uses deep neural networks in order to perform the reduction, has been introduced. The DNN has as a benefit that it directly optimizes the approximation error (squared Frobenius norm in this case) and directly gives the matrices of the reduced LPV model whereas for the KPCA and AE method an additional optimization step is required. Furthermore, as suggested in this paper, using a deep encoding layer with ReLU functions in the DNN method allows for better approximation of more complex scheduling maps while still being able to compute the output of the encoding layer fast and efficiently.
From the results on the two-link robotic arm example it can also be concluded that the developed DNN method results in a improved representation (of the matrix variations) of the original model (in terms of average Frobenius norm squared) and results in improved tracking performance when the (reduced) LPV model is used for synthesis compared to the current methods.
While an affine scheduling dependency is considered in this paper, many nonlinear models can be embedded in LPV models with less conservativeness by considering a different scheduling dependency, such as a rational dependency. Hence, for future research, the possibility of applying SDR to LPV models with other scheduling dependencies (such as rational) needs to be investigated. Moreover, the influence of different approximation error cost functions, such as $\mathcal{H}_\infty/\mathcal{H}_2$ based error in terms of the local frequency responses, needs to be investigated as currently only Frobenius norm based cost functions on the matrix variations have been used.

%

\bibliographystyle{IEEEtran}
\bibliography{references}


\end{document}